\documentclass[final,5p,times,twocolumn]{elsarticle}

\usepackage{graphics}
\usepackage{amssymb}
\usepackage{comment}

\def\Journal#1#2#3#4{{#1} {\bf #2} (#4) #3}
\def\APJS{Astrophys. J. Suppl}

\def\JCAP{JCAP}
\def\JHEP{JHEP}

\def\NPB{Nucl. Phys. B}

\def\PLB{{Phys. Lett.} B}
\def\PREP{Phys. Rep.}

\def\PRL{Phys. Rev. Lett.}
\def\PRD{Phys. Rev. D}

\def\PTP{Prog. Theor. Phys.}
\def\RMP{Rev. Mod. Phys.}

\journal{Physics Letters B}

\begin{document}

\begin{frontmatter}



\title{CP violation in modified bipair neutrino mixing and leptogenesis}


\author{Jun Iizuka}
\author{Yoshikazu Kaneko}
\author{Teruyuki Kitabayashi\corref{cor1}}
\ead{teruyuki@keyaki.cc.u-tokai.ac.jp}
\author{Naoto Koizumi}
\author{Masaki Yasu\`{e}}
\ead{yasue@keyaki.cc.u-tokai.ac.jp}
\cortext[cor1]{Corresponding author}
\address{Department of Physics, Tokai University, 4-1-1 Kitakaname, Hiratsuka, Kanagawa, 259-1292, Japan}

\begin{abstract}
We study effects of CP violation in a modified bipair neutrino mixing scheme predicting $\sin^2\theta_{23}$ near both 0.4 and 0.6 currently consistent with experimentally allowed values. The source of CP violation is supplied by charged lepton mixing accompanied by a single phase, whose mixing size is assumed to be less than that of the Wolfenstein parameter for the quark mixing. Including results of leptogenesis, which is based on the minimal seesaw model, we obtain the allowed region of CP-violating Dirac and Majorana phases, which provides the observed baryon asymmetry of the universe in the case of the Dirac neutrino mass matrix subject to one zero texture.
\end{abstract}

\begin{keyword}
Modified bipair neutrino mixing \sep CP-violating phases \sep Leptogenesis

\end{keyword}

\end{frontmatter}

% \linenumbers


%---------------------------------------------
%\section{Introduction}
{\bf Introduction: }
%---------------------------------------------
One of the important and unsolved problems in neutrino physics is understanding of the CP properties of neutrinos, where CP violation may occur. For three flavor neutrinos, CP violation is induced by one CP-violating Dirac phase $\delta_{CP}$ and two CP-violating Majorana phases $\alpha_2$ and $\alpha_3$.  The CP-violating  Majorana phases completely disappear from the oscillation probabilities and cannot be measured by quite familiar oscillation experiments \cite{Mohapatra_Pal}. Although the CP-violating Majorana phases can enter in processes of neutrinoless double beta decay, the detection of the Majorana CP violation has not been succeeded \cite{DoubleBeta}. On the other hand, in the leptogenesis scenario \cite{leptogenesis}, the baryon-photon ratio in the universe \cite{etaB} is generated if CP-violating Majorana phases exist. 

Recently, T.K. and M.Y. have studied the correlation between CP-violating Dirac and Majorana phases in the modified bipair neutrino mixing scheme without consideration of the leptogenesis \cite{KitabayashiYasue2013}. The allowed region of these CP-violating phases, which is consistent with the observed neutrino mixing angles has been determined.

In this Letter, including leptogenesis analysis, we re-examine the allowed region of CP-violating Dirac and Majorana phases in a modified bipair neutrino mixing scheme. 

%---------------------------------------------
%\section{Modified bipair neutrino mixing}
{\bf Modified bipair neutrino mixing: }
%---------------------------------------------
The neutrinos have tiny masses $m_i$ $(i=1,2,3)$ and exhibit non-zero mixing angles $\theta_{ij}$ $(i,j=1,2,3)$ . There are theoretical discussions that predict these mixing angles in literatures \cite{reviewObMixings}. The bipair neutrino mixing scheme is one of such mixing textures. The original bipair neutrino mixing \cite{KitabayashiYasue2011} is described by a mixing matrix which is equipped with two pairs of identical magnitudes of matrix elements to be denoted by $U^0_{ij}$ ($i.j$=1,2,3). There are two cases of the bipair texture. The case 1 of the bipair neutrino mixing ($\vert U^0_{12} \vert = \vert U^0_{32} \vert$ and $\vert U^0_{22} \vert = \vert U^0_{23} \vert$) is parameterized by $U^0_{BP1}$ and the case 2 of the bipair neutrino mixing ($\vert U^0_{12} \vert = \vert U^0_{22} \vert$ and $\vert U^0_{32} \vert = \vert U^0_{33} \vert$) is parameterized by  $U^0_{BP2}$ as follows:
\begin{equation}
U^0_{BP1}=\left(
  \begin{array}{ccc}
    c      & s   & 0 \\
    -t^2  & t   & t \\
    st     & -s  & t/c 
  \end{array}
\right),
U^0_{BP2}=\left(
  \begin{array}{ccc}
     c      & s   & 0 \\
     -st   & s   & t/c  \\
      t^2  & -t  & t 
  \end{array}
\right),
\label{Eq:UnuBP}
\end{equation}
where $c=\cos\theta_{12}$, $s=\sin\theta_{12}$ and $t=\tan\theta_{12}$. The mixing angles are predicted to be $\sin^2\theta_{12} = 1 - 1/\sqrt{2} = 0.293$, $\sin^2\theta_{23} = \sqrt{2} - 1 = 0.414$ and $\sin^2\theta_{13} = 0$ in the case 1 while the mixing angles are predicted to be $\sin^2\theta_{12} = 1 - 1/\sqrt{2} = 0.293$, $\sin^2\theta_{23} = 2 - \sqrt{2} = 0.586$ and $\sin^2\theta_{13} = 0$ in the case 2. It should be noted that the bipair neutrino mixing scheme does predict the larger value of $\theta_{23}$ consistent with its experimentally allowed value around 0.6 \cite{chargedLeptonContributions}.

The original bipair neutrino mixing predicts $\sin^2\theta_{13} = 0$ which is inconsistent with the observation. It is expected that additional contributions to the mixing angles are produced by charged lepton contributions  \cite{chargedLeptonContributions} if some of the non-diagonal matrix elements of charged lepton mass matrix are nonzero so that the reactor mixing angle can be shifted to lie in the allowed region. 

We follow the modification scheme in the Ref.\cite{KitabayashiYasue2013}. The modified bipair neutrino mixing $U$ is given by $U= U_\ell^\dagger U_\nu$, where $U_{\ell}$ (as well as $U_R$ to be used later) and $U_\nu$, respectively, arise from the diagonalization of the charged lepton mass matrix $M_\ell$ and of the neutrino mass matrix $M_\nu$. The neutrino mixing matrix $U_\nu$ can be further parameterized to be $U_\nu =  P U^0_\nu$ with $U^0_\nu = {\tilde U}^0_\nu K$, where $P$ is defined by $P = {\rm diag}(1, e^{i\phi_2}, e^{i\phi_3})$ with phases $\phi_2$ and $\phi_3$ and, similarly, $K$ = ${\rm diag}(1, e^{i\rho_2}, e^{i\rho_3})$. As a result, ${\tilde U}^0_\nu$ contains a CP-violating phase of the Dirac type. The lepton mass matrices satisfy the relations of $M_\ell = U_\ell M_\ell^{diag} U_R^\dagger$ and $M_\nu = U_\nu M_\nu^{diag} U_\nu^T$, where $M_\ell^{diag}$ and $M_\nu^{diag}$ are, respectively, the diagonal mass matrix of charged leptons and of neutrinos. Flavor neutrinos, therefore, form a mass matrix ${M_f}$ given by ${M_f} = U_\ell^T{M_\nu }U_\ell$. In order to estimate the charged lepton contributions to the mixing angles, we take the neutrino mixing matrix to be either $U^0_\nu = U_{BP1}^0$ or $U^0_\nu = U_{BP2}^0$.

The charged lepton mixing matrix $U_\ell$ can be parameterized by three mixing angles and one phase $\delta_\ell$. To be more specific, we adapt the Cabibbo-Kobayashi-Maskawa like parameterization of the matrix $U_\ell$ \cite{chargedLeptonContributions,Pascoli2003}:
\begin{equation}
U_\ell= 
\left( {\begin{array}{*{20}{c}}
{1 - \frac{{\epsilon _{12}^2 + \epsilon _{13}^2}}{2}}&{{\epsilon _{12}}}&{{e^{ - i\delta_\ell }}{\epsilon _{13}}}\\
{ - {\epsilon _{12}} - {\epsilon _{23}}{\epsilon _{13}}{e^{i\delta_\ell }}}&{1 - \frac{{\epsilon _{12}^2 + \epsilon _{23}^2}}{2}}&{{\epsilon _{23}}}\\
{{\epsilon _{12}}{\epsilon _{23}} - {e^{i\delta_\ell }}{\epsilon _{13}}}&{ - {\epsilon _{23}} - {\epsilon _{12}}{\epsilon _{13}}{e^{i\delta_\ell }}}&{1 - \frac{{\epsilon _{23}^2 + \epsilon _{13}^2}}{2}}
\end{array}} \right),
\label{Eq:generalUell-approximated}
\end{equation}
where $\epsilon_{ij}$ $(i,j=1,2,3)$ is small parameter  having magnitude of the order of Wolfenstein parameter $\sim 0.227$ for the quark mixing or less. The most simple case will be obtained in the case of $\epsilon_{12}=\epsilon_{23}=\epsilon_{13}=\epsilon$. Hereafter,  we consider this simple case. We note that the unitarity problem with Eq.(\ref{Eq:generalUell-approximated}) has been discussed in Ref.\cite{KitabayashiYasue2013}. 

It is known that CP-violating Dirac phase $\delta_{CP}$ is determined by the Jarlskog invariant \cite{Jarlskog1985}
$\sin\delta_{CP} = {\rm Im} \left(U_{11}U_{22}U_{12}^\ast U_{21}^\ast \right) / (c_{12}c_{23}c_{13}^2s_{12}s_{23}s_{13})$.
One can also find CP-violating Majorana phases $\alpha_2, \alpha_3$ to be \cite{KitabayashiYasue2013}:
$\alpha_2= {\rm arg} \left(U_{12}U_{11}^\ast\right)$ and $\alpha_3= {\rm arg} \left(U_{13}U_{11}^\ast\right) + \delta_{CP}$.

%%----------------------------------------------------------------------------------
%\section{Leptogenesis}
{\bf Leptogenesis: }
%%----------------------------------------------------------------------------------
The tiny neutrino masses could be explained by the so-called seesaw mechanism \cite{Seesaw}. In the minimal seesaw model \cite{minimalSeesaw}, we introduce two right-handed neutrinos $N_1$ and $N_2$ into the standard model and obtain a light neutrino mass matrix by the relation of $M_f = -m_D M_R^{-1} m_D^T$, where $m_D$ is a Dirac neutrino mass matrix and $M_R$ is a right-handed neutrino mass matrix. 

The baryon-photon ratio in the universe $\eta_B$ can be generated by leptogenesis within the framework of the minimal seesaw model in collaboration with the heavy neutrinos and Higgs scalar. The recipes to predict $\eta_B$ are given in Ref. \cite{FlavoredLeptogenesis} as follows:

(I)We assume that the mass matrix of the right-handed neutrinos is diagonal and real: $M_R = {\rm diag}(M_1, M_2)$ and that no phases are associated with the heavy neutrinos. The CP asymmetry from the decay of $N_1$ (we assume $M_1 \ll M_2$) is given by the flavor-dependent $\lambda_\alpha$ $(\alpha = e, \mu, \tau)$:
\begin{eqnarray}
\lambda_\alpha= \frac{1}{8\pi v^2}\frac{{\rm Im}[(m_D^\dagger)_{1\alpha}(m_D)_{\alpha 2}(m_D^\dagger m_D)_{12}]}{(m_D^\dagger m_D)_{11}}f\left(\frac{M_2}{M_1}\right),
\end{eqnarray}
where $v \simeq 174$ GeV and 
\begin{eqnarray}
f(x) = x\left( 1-(1+x^2)\ln\left(\frac{1+x^2}{x^2}\right) + \frac{1}{1-x^2} \right).
\end{eqnarray}
(I\hspace{-.1em}I) The washout effect on $\lambda_\alpha$ in the expanding universe is controlled by $\eta(m_{1eff}^\alpha)$, where 
\begin{eqnarray}
\eta(x) = \left( \frac{8.25\times 10^{-3} {\rm eV}}{x} + \left( \frac{x}{2\times 10^{-4} {\rm eV}} \right)^{1.16} \right)^{-1},
\label{Eq:eta}
\end{eqnarray}
and $m_{1eff}^\alpha = (m_D^\dagger)_{1\alpha} (m_D)_{\alpha 1}/M_1$.

(I\hspace{-.1em}I\hspace{-.1em}I) For our adopted range of $M_1$, see below,  the lepton number in the co-moving volume is calculated to be
\begin{eqnarray}
Y_L &\simeq& \frac{12}{37 g_\ast}\left [(\lambda_e+\lambda_\mu)\eta\left(\frac{417}{589}(\vert a_1\vert^2 + \vert a_2\vert^2 )\right) \right.
 \nonumber \\
&&\left.+ \lambda_\tau \eta\left(\frac{390}{589} \vert a_3 \vert^2\right)  \right],
\label{Eq:YL}
\end{eqnarray}
where we take the Dirac neutrino mass matrix in terms of 6 parameters $a_1, a_2, a_3, b_1, b_2, b_3$ as
\begin{eqnarray}
m_D = 
\left(
  \begin{array}{cc}
    a_1  & b_1   \\
    a_2  & b_2   \\
    a_3  & b_3   \\
  \end{array}
\right).
\end{eqnarray}

(I\hspace{-.1em}V) The effective number of the relativistic degree of freedom $g_\ast$ is calculated as \cite{KolbTurner1990}
\begin{eqnarray}
g_\ast = \sum_{i=bosons} g_i\left(\frac{T_i}{T}\right)^4 + \frac{7}{8} \sum_{i=fermions} g_i\left(\frac
{T_i}{T}\right)^4,
\end{eqnarray}
where $T$ is thermal equilibrium temperature of the universe, $T_i$ and $g_i$ are temperature and number of internal degrees of freedom of the relativistic particle species $i$. We assume that all the particle species in the standard model were relativistic when the leptonic CP-asymmetry was generated by the decay process of the lightest right-handed neutrino $N_1$, while $N_1$ was heavy enough to be non-relativistic itself. We take $g_\ast=106.75$.

(V) The baryon number in the co-moving volume is $Y_B=-0.54Y_L$. Finally, the baryon-photon ratio is estimated to be $\eta_B=7.04Y_B$.

According to the condition of $\det(M_f)=0$, at least one of the neutrino mass eigenvalues $(m_1,m_2,m_3)$ must be zero in the minimal seesaw model \cite{minimalSeesaw}. We obtain the two types of  hierarchical neutrino mass spectrum in the minimal seesaw model. One is the normal mass hierarchy $(m_1,m_2,m_3)=\left(0,\sqrt{\Delta m_\odot^2},\sqrt{\Delta m_{atm}^2}\right)$ and the other is the inverted mass hierarchy $(m_1,m_2,m_3)=\left(\sqrt{\Delta m_{atm}^2},\sqrt{\Delta m_{atm}^2-\Delta m_\odot^2},0\right)$, where the squared mass differences of solar and atmospheric neutrinos, respectively,  defined by $\Delta m_{\odot}^2$ $\equiv$ $m_2^2-m_1^2$ and $\Delta m_{atm}^2$ $\equiv$ $m_3^2-m_1^2$. We use $\Delta m_{\odot}^2 =7.50 \times 10^{-5} {\rm eV}^2$ and $\Delta m_{atm}^2  = 2.473 \times 10^{-3} {\rm eV}^2$\cite{Gonzalez-Garcia2012}. Since we would like to discuss effects of the CP-violating Dirac phase on the creation of lepton number $Y_L$, we may consider the renormalization effects that modify the magnitude of when is promoted to $Y_L$. It has been discussed that the renormalization effect is rather insignificant for neutrinos in the normal mass hierarchy \cite{renormalization_effects}, where we reside now. Moreover, because of the mass spectrum of the minimal seesaw, only a single Majorana phase is physically relevant. For the rest of paper, we use $\alpha=\alpha_3-\alpha_2$ to denote the physically relevant CP-violating Majorana phase. 
%
%--------------------------------------------------------------------
%--------------------------------------------------------------------
\begin{figure}[!t]
\begin{center}
 \includegraphics{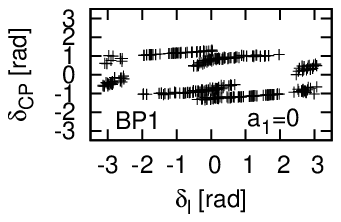}
 \includegraphics{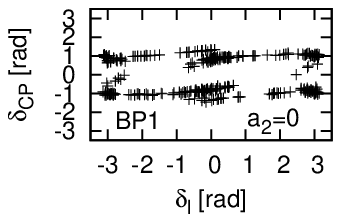}
 \includegraphics{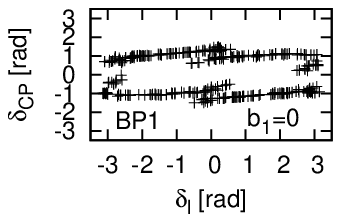}
 \includegraphics{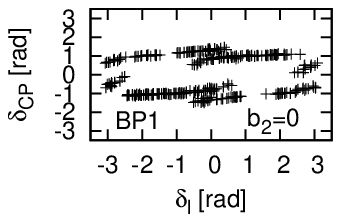}
 \includegraphics{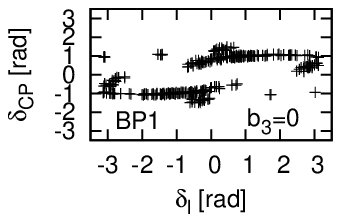}
  \caption{\label{fig:fig1} The mutual dependence of $(\delta_{CP},\delta_\ell)$ in the case 1.}
\end{center}
\end{figure}
%--------------------------------------------------------------------
\begin{figure}[!h]
\begin{center}
 \includegraphics{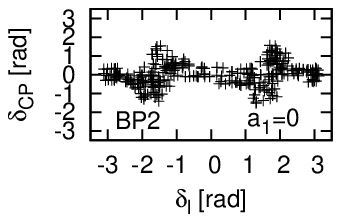}
 \includegraphics{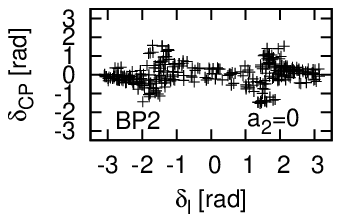}
 \includegraphics{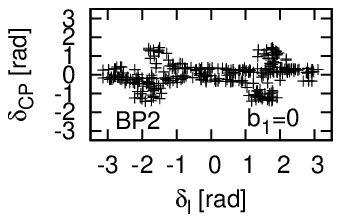}
 \includegraphics{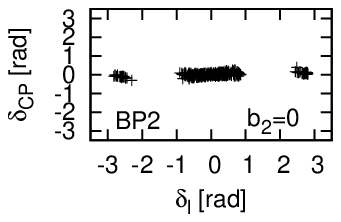}
 \includegraphics{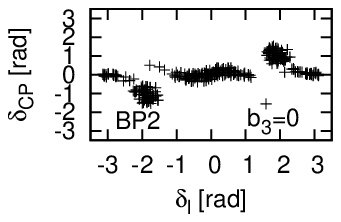}
  \caption{\label{fig:fig2} Same as Figure \ref{fig:fig1} but in the case 2.}
\end{center}
\end{figure}
%--------------------------------------------------------------------
%--------------------------------------------------------------------
\begin{figure}[!t]
\begin{center}
 \includegraphics{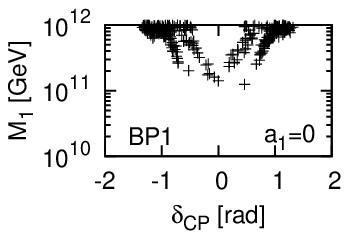}
 \includegraphics{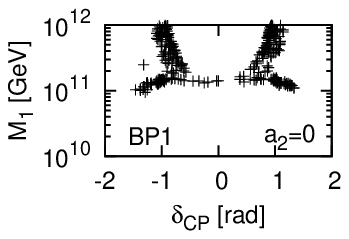}
 \includegraphics{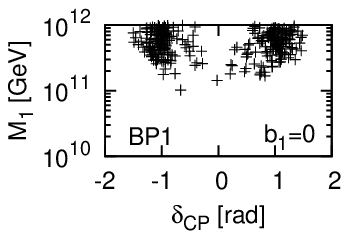}
 \includegraphics{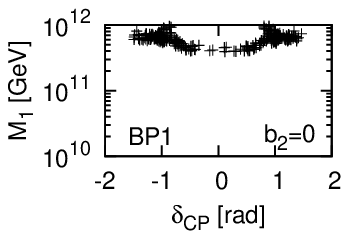}
 \includegraphics{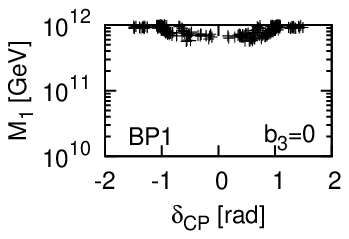}
  \caption{\label{fig:fig3} The mutual dependence of $(M_1,\delta_{CP})$ in the case 1.}
\end{center}
\end{figure}
%--------------------------------------------------------------------
\begin{figure}[!h]
\begin{center}
 \includegraphics{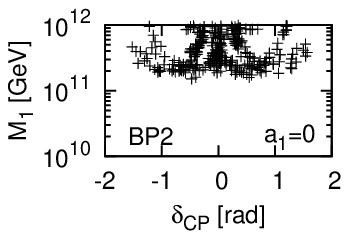}
 \includegraphics{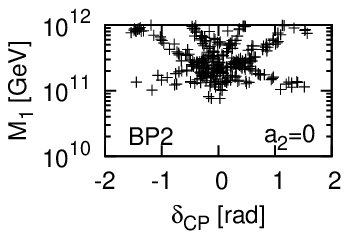}
 \includegraphics{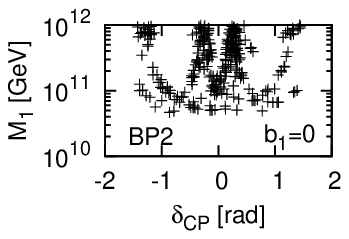}
 \includegraphics{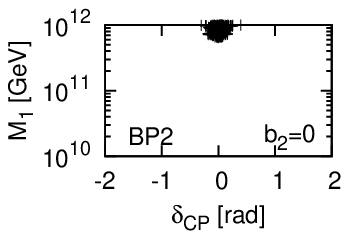}
 \includegraphics{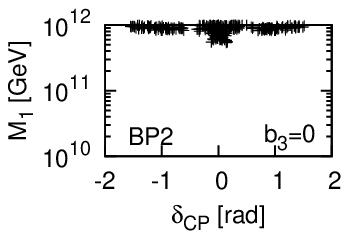}
  \caption{\label{fig:fig4} Same as Figure \ref{fig:fig3} but in the case 2.}
\end{center}
\end{figure}
%--------------------------------------------------------------------
%--------------------------------------------------------------------
\begin{figure}[!t]
\begin{center}
 \includegraphics{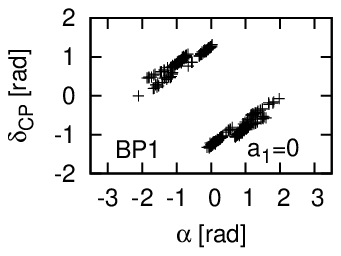}
 \includegraphics{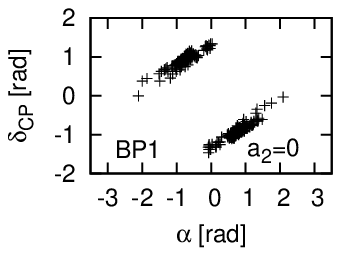}
 \includegraphics{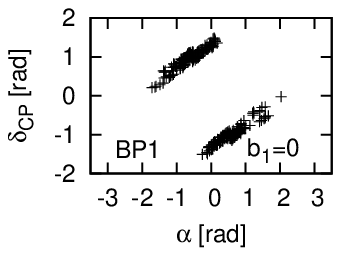}
 \includegraphics{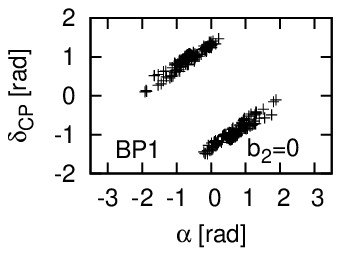}
 \includegraphics{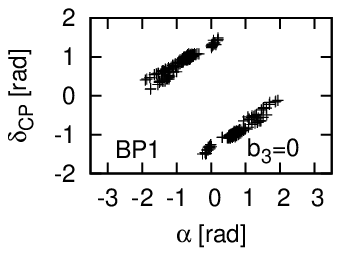}
  \caption{\label{fig:fig5} The mutual dependence of $(\delta_{CP},\alpha)$ in the case 1.}
\end{center}
\end{figure}
%--------------------------------------------------------------------
\begin{figure}[!h]
\begin{center}
 \includegraphics{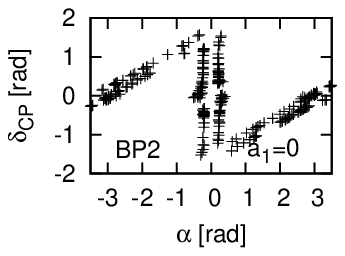}
 \includegraphics{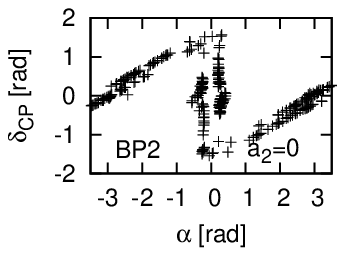}
 \includegraphics{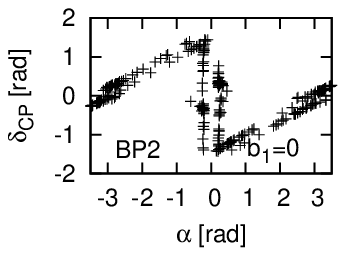}
 \includegraphics{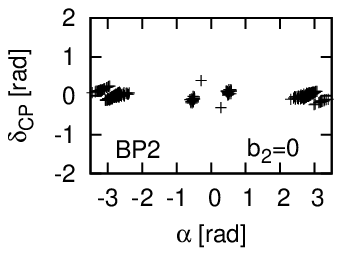}
 \includegraphics{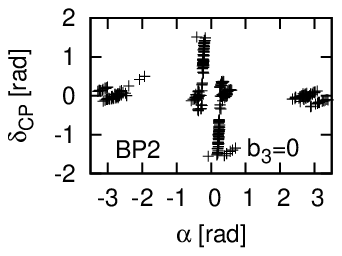}
  \caption{\label{fig:fig6} Same as Figure \ref{fig:fig5} but in the case 2.}
\end{center}
\end{figure}
%--------------------------------------------------------------------
%--------------------------------------------------------------------

%\section{Numerical calculation}
{\bf Sizes of CP-violating phases: }
To estimate sizes of CP-violating phases, we perform the numerical calculation with the following setups:

(I) The flavor neutrino mass matrix $M_f$ described by six neutrino masses contains five complex parameters because of the condition of $\textrm{det}(M_f)=0$. If the Dirac mass matrix $m_D$, which contains six complex parameters, has so-called one zero texture \cite{texturezero}, we can analytically express the Dirac mass matrix elements in $m_D$ in terms of the light neutrino masses in $M_f$. We assume that either one of $a_1, a_2, b_1, b_2, b_3$ is equal to zero.  The case with $a_3=0$ should be omitted to avoid divergence arising from Eq.(\ref{Eq:eta}) and Eq.(\ref{Eq:YL}).

(I\hspace{-.1em}I) For an input mass parameter $M_1$, we consider the case of $M_1 \lesssim 10^{12}$ GeV, where flavor effects are relevant \cite{FlavoredLeptogenesis}. The heavier neutrino mass $M_2$ is fixed to be $M_2 = 10^3M_1$. Other input parameters are taken to vary in the range of $0\leq \epsilon \leq 0.227$ and $-\pi \leq \phi_2, \phi_3, \delta_\ell \leq \pi$.

(I\hspace{-.1em}I\hspace{-.1em}I) Remaining input parameters are the 1$\sigma$ data of neutrino mixing angles \cite{Gonzalez-Garcia2012}, e.g., $\sin^2 \theta_{12} = 0.302^{+0.013}_{-0.012}$ and $\sin^2 \theta_{13} = 0.0227^{+0.0023}_{-0.0024}$ as well as $\sin^2 \theta_{23} = 0.413^{+0.037}_{-0.025}$ for the case 1 and $\sin^2 \theta_{23} = 0.594^{+0.021}_{-0.022}$ for the case 2 together with the observed baryon-photon ratio $\eta_B = 6.160^{+0.153}_{-0.156} \times 10^{-10}$  \cite{etaB}.

(I\hspace{-.1em}V) In order to ensure the thermal leptogenesis to be the source of the baryon asymmetry in the universe, the reheating 
temperature after inflation must have been greater than the mass scale of the lightest right-handed neutrino \cite{SUSYproblem}. Hence the lower bound on the reheating temperature must be greater than $\sim 10^{10}$ GeV. However, this high reheating temperature is not suitable for 
supersymmetric (SUSY) theories because it may lead to an overproduction of light supersymmetric 
particles, such a gravitino after inflation \cite{SUSYproblem}. We are not considering this problem 
here and are limiting discussion on non-SUSY cases.

Figure \ref{fig:fig1} for the case 1 (BP1) and Figure \ref{fig:fig2} for the case 2 (BP2) show how $\delta_{CP}$ in the neutrino mixing is correlated with $\delta_\ell$ in the charged-lepton mixing. In these figures, $a_1=0$, $a_2=0$ and so on described in the bottom right corner specify the position of the zero element of the texture one zero Dirac mass matrix $m_D$. The almost entire range of $\delta_\ell$ provides $\delta_{CP}$, which is constrained to yield the observed $\eta_B$.  Roughly speaking, we find that, for the wider region of the allowed $\delta_{CP}$, the region of $\delta_{CP}$ in the case 1 is complementary to the one in the case 2. This trend cannot be observed in other figures, where each figure is a function of $\delta_{CP}$.

Figure \ref{fig:fig3} shows the mutual dependence of the mass of right-handed neutrino $M_1$ and CP-violating phases $\delta_{CP}$ in the case 1 of the modified bipair neutrino mixing. Figures \ref{fig:fig4} shows the same as Figures \ref{fig:fig3} but in the case 2. We find that $M_1$ is constrained as follows:
\begin{itemize}
\item $M_1 \lesssim 1-5 \times 10^{11}$ GeV in the case 1,
\item $M_1 \lesssim 0.5-7 \times 10^{11}$ GeV in the case 2,
\item More stringent constraint of $M_1 \gtrsim {\rm several} \times 10^{11}$ GeV near its maximal value of $10^{12}$ GeV is found for $b_2=0$ of $b_3=0$.
\end{itemize}

Figures \ref{fig:fig5} shows the mutual dependence of the CP-violating phases $\delta_{CP}$ and $\alpha$ in the case 1 of the modified bipair neutrino mixing. Figures \ref{fig:fig6} shows the same as Figures \ref{fig:fig5} but in the case 2. We see overall behavior of the mutual dependence of $\delta_{CP}$ [rad] and $\alpha$ [rad] as follows:
\begin{itemize}
\item in the case 1 for the all cases;
\begin{itemize}
\item $\vert \alpha\vert$ approaches to its maximal value of $\pi/2$ as $\vert \delta_{CP}\vert$ approaches to its minimal value around 0,
\item $\vert \alpha\vert$ approaches to its minimal value of 0 as $\vert \delta_{CP}\vert$ approaches to its maximal value of $\pi/2$,
\item roughly speaking, $\delta_{CP}$ is scattered around the straight line of $\vert \delta_{CP} \vert= \pi(\vert \alpha \vert - 2)/4$,
\end{itemize}
\item in the case 2 for $a_1=0$, $a_2=0$ or $b_1=0$;
\begin{itemize}
\item $\vert \alpha\vert$ approaches to its maximal value of $\pi$ as $\vert \delta_{CP}\vert$ approaches to its minimal value of 0,
\item $\vert \alpha_\vert$ approaches to its minimal value of 0 as $\vert \delta_{CP}\vert$ approaches to its maximal value of $\pi/2$,
\item roughly speaking, $\delta_{CP}$ is scattered around the straight line of $\vert \delta_{CP} \vert = (\vert \alpha \vert - \pi)/2$ except for $\alpha=0$,
\item $\delta_{CP}$ takes any allowed values for $\alpha \sim 0$,
\end{itemize}
\item in the case 2 for $b_2=0$;
\begin{itemize}
\item $\delta_{CP} \sim 0$,
\item $\vert \alpha \vert \sim 0.5$ or $3\pi/4 \lesssim \vert \alpha \vert \lesssim \pi$,
\end{itemize}
\item in the case 2 for $b_3=0$;
\begin{itemize}
\item $\delta_{CP} \sim 0$ or $\delta_{CP}$ takes any allowed values for $\alpha \sim 0$,
\item $\vert \alpha \vert \lesssim 0.5$ or $3\pi/4 \lesssim \vert \alpha \vert \lesssim \pi$.
\end{itemize}
\end{itemize}
%
%%----------------------------------------------------------------------------------
%\section{Summary}
{\bf Summary and Discussions:}
%%---------------------------------------------------------------------------------- 
We have assumed that CP-violating phases are induced by charged lepton mixing including the phase $\delta_\ell$ accompanied by its mixing parameter $\epsilon$, whose size is taken to be less than that of the Wolfenstein parameter for the quark mixing. The Dirac mass matrix is limited to one zero texture and $M_1$ is constrained to be: $M_1 \lesssim 10^{12}$ GeV ($\ll M_2)$, where the flavor effects are relevant for the leptogenesis.

The following predictions are obtained:
\begin{enumerate}
\item
The larger CP-violating Dirac phase around $\vert \delta_{CP}\vert \sim \pi/2$ favors for the suppressed CP-violating Majorana phase $\alpha$ around $\alpha \sim 0$ except for $b_2=0$ in the case 2, where CP-violating Dirac phase should be small in this case.
\item
The smaller CP-violating Dirac phase around $\vert \delta_{CP}\vert \sim 0$ favors for the larger CP-violating Majorana phase $\alpha$ around $\vert\alpha\vert \sim \pi/2$ in the case 1 or around $\alpha \sim 0,\pm\pi$ in the case 2.
\item $\delta_{CP}$ is scattered around the straight line of $\vert \delta_{CP} \vert= \pi(\vert \alpha \vert - 2)/4$ in the case 1 or $\vert \delta_{CP} \vert = (\vert \alpha \vert - \pi)/2$ except for $\alpha=0$ in the case 2 ($\delta_{CP}$ takes any allowed values for $\alpha \sim 0$ in this case).
\item The most stringent constraints on the CP-violating phases are obtained in the case 2 with $b_2=0$ as $\delta_{CP} \sim 0$ and $\vert \alpha \vert \sim 0.5$ or $3\pi/4 \lesssim \vert \alpha \vert \lesssim \pi$.
\end{enumerate}
Furthermore, if the CP-violating Majorana phase is suppressed, the case 1 of the smaller values of $\sin^2\theta_{23}$ needs $\delta_{CP}\sim \pi/2$ while the case 2 of the larger values of $\sin^2\theta_{23}$ except for the $b_2=0$ case allows the nonvanishing $\delta_{CP}$ covering a wide range of $-\pi/2\lesssim\delta_{CP}\lesssim\pi/2$. For $b_2=0$ in the case 2, $\delta_{CP}$ is also suppressed and the larger $M_1$ near its maximally allowed value is thus required.
%%--------------------------------
%% Acknowledgments
%%--------------------------------
%\vspace{10pt}
%{\bf Acknowledgments:}


%%appendix

%% References


\bibliographystyle{elsarticle-num}
\bibliography{<your-bib-database>}



\end{document}